# Ion irradiation and biomolecular radiation damage Ⅱ. Indirect effect

Wei WANG [*†‡] Zengliang YU[‡] Wenhui SU [†]

[†]CCMST, Academy of Fundamental and Interdisciplinary Sciences, Harbin Institute of Technology, Harbin 150080, China；
[‡] Key Lab of Ion Beam Bioengineering, Institute of Plasma Physics, Chinese Academy of Sciences, P. O. Box 1126, Hefei, 230031, China

E-mail: wwang_ol@hit.edu.cn

## 1. Introduction

Ions exist everywhere in nature – in cosmic radiation, air, water, and living bodies. The interstellar space, the ionospheres of planets are all full of ions. Near the surface of the Earth, there are also a large amount of low energy particles generated from aurora, lightning, volcanic eruption and emission of radioactive elements in the crust. Will you, nill you, all life on the Earth expose themselves to an "ionic atmosphere" and these low energy ions inevitably bring influence to bear on all life. Otherwise，energetic ions with energy rang from keV to MeV, are ubiquitous in extraterrestrial space, such as low-energy cosmic rays, magnetospheric ions, solar flares, and solar wind particles. These charged particles whiz around the Earth and other celestial bodies in huge numbers and constantly bombard Earth's upper atmosphere. Along with the extension of human activities to the space, their biological effect on human beings should also been paid urgent attention.

In the mid-1980's, we introduce 30keV nitrogen ions to dry rice seeds. When the implantation dose was sufficiently high, yellow stripes were seen on the leaves of rice plants grown from the seeds, and these yellow characteristics could be stably inherited to later generations[1]. This is the beginning of low energy ion biology.

Since late 1980's, we have began embarking on ion-beam bioengineering research (e.g., ion implantation to improve corps and microbes, ion beam mediated gene transfer) and great achievements have been made[2-5]. Definitely, all these observable events are correlative with chemical changes. For a more indepth study on this subject, we need to uncover what has happened to those organisms at the molecular level and what chemical changes have occurred to the biomolecules .

As we all know, Ion impact-induced fragmentation of molecules is a process of fundamental importance in many areas of science and technology, for example, radiation damage to biological tissue. Factors related to ion implantation induced biological effects, as distinguished from those due to radiation, include not only energy deposition but also mass deposition and charge exchange[1]. When its energy drops down to the energy range of chemical reactions, the implanted atom may take part in chemical reactions, directly or indirectly, with target atoms and become either an interstitial or a substitutional atom. The target atoms may rearrange and recombine to form a new compound. Accordingly, ion implantation may bring damage to biomolecules and organisms.

Biomolecular damage (e.g., amino acids damage, nucleobase damage) is the base of DNA damage[6-8], chromosome aberration[5], and other biological effects of ion beam. It has been reported that damage of genome in a living cell by ionizing radiation is about one-third direct and two-thirds indirect[9,10]. The former which has been introduced in our last paper[11], concerns direct energy deposition and ionizing reactions in the biomolecules; the latter results from radiation induced reactive species (mainly radicals) in the medium (mainly water) surrounding the biomolecules. In this review, a short description of ion implantation induced radical formation in water is presented. Then we summarize the aqueous radical reaction chemistry of DNA, protein and their components, followed by a brief introduction of biomolecular damage induced by secondary particles (ions and electron). Some downstream biological effects are also discussed.

## 2. Ion beam radiation chemistry of liquid water

We always used to say that small biomolecules (e.g., amino acids) are the fundamental building blocks of life. Water, however, is also an important constituent of living matter (65-90% for different organisms). Radiation damage of biological tissues during radiotherapy or exposure to nature radiation is therefore closely connected to water radiolysis. In view of this, information on the ionization of water molecules is fundamental in studies of interaction of energetic particles with biological material.

The history of radiation chemistry of water with energetic ions is only a few years shy of the discovery of natural radioactivity, and now the radiolysis of water is a phenomenon known for already one century and well understood[12]. Curie and Debierne observed that radium emanations decompose water into gases which were demonstrated to be the hydrogen and oxygen[13]. Further studies with deaerated water led to the identification of the molecular yield consisting of $H_2$ and $H_2O_2$ which was attributed to the recombination reactions of primary H· and OH· radicals. The slow decomposition of $H_2O_2$ explains the formation of $O_2$. In the 1950's Platzmann postulated that free electrons can be stabilized in water based on theoretical considerations, i.e., the hydrated electron ($e_{aq}^-$). This theory has been demonstrated in the 1960's[14].

In the first half of the 20th century, α-emitters such as polonium and curium were solely used for ion beam irradiation. Nowadays, in addition to traditional sources, various kinds of ion accelerators are at an industrial stage, such as cyclotrons or synchrotrons for high-energy ion beam, Van de Graff or single-ended accelerators for medium energy, and ion implanters for low energy. A large number of facilities are ready for making studies employing ion beam. Many of recent studies with ion beam are more closely associated with material science and life science due to their industrial and medical applications. However, a complete understanding of track processes in radiation chemistry and radiation biology will only come about by the experimental examination of basic process coupled with theoretical and kinetic models.

Water radiolysis by ion implantation received a lot of interest both experimentally[15-18] and theoretically[19-22]. A variety of reactive species possibly exist in the radiation track of the charged particle [21, 23].

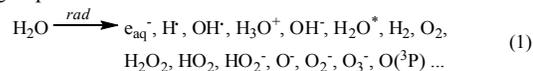

$$H_2O \xrightarrow{rad} e_{aq}^-, H\cdot, OH\cdot, H_3O^+, OH^-, H_2O^*, H_2, O_2,$$
$$H_2O_2, HO_2, HO_2^-, O^-, O_2^-, O_3^-, O(^3P) \ldots \qquad (1)$$

All energetic charged particles produce the same ubiquitous and radiation-biologically important transient, H·, OH·, and $e_{aq}^-$, each with a nonhomogeneous distribution along the penetration range. Low energy ion does not make an exception[24]. G-value and LET (Linear Energy Transfer) can be used to characterize the trace amount of chemical yields and the energy transferred to medium as an ionizing particle travels through it, respectively. G-values of the products in water irradiated with ion beams of different energies are shown in Table 1[25]. The studies on the radiolysis of water and aqueous solutions showed that with increasing particle LET there was an increase in the yields of molecular products such as $H_2$ and $H_2O_2$ and a corresponding decrease in the yields of radicals such as ·H and ·OH[12,26].

Table 1. G values for various species at $10^{-7}$s for protons of several energies and for Alpha particles of the same velocities (ref. 25)

| Species | Protons (MeV) | | | | Alpha Particles (MeV) | | | |
|---|---|---|---|---|---|---|---|---|
| | 1 | 2 | 5 | 10 | 4 | 8 | 20 | 40 |
| OH | 1.05 | 1.44 | 2.00 | 2.49 | 0.35 | 0.66 | 1.15 | 1.54 |
| H | 1.37 | 1.53 | 1.66 | 1.81 | 0.79 | 1.03 | 1.33 | 1.57 |
| $e_{aq}^-$ | 0.19 | 0.40 | 0.83 | 1.19 | 0.02 | 0.08 | 0.25 | 0.46 |
| $H_2$ | 1.22 | 1.13 | 1.02 | 0.93 | 1.41 | 1.32 | 1.19 | 1.10 |
| $H_2O_2$ | 1.48 | 1.37 | 1.27 | 1.18 | 1.64 | 1.54 | 1.41 | 1.33 |

In many biological experiments the low energy particle beam is completely stopped in the sample and it loses the whole energy in the medium. Mass deposition will also happen to non-inert gas ions (e.g., nitrogen ions) in ion implanted water, as well as energy deposition. The Hefei group[27,28] has developed a heuristic model for low energy $N^+$ ion implantation. They found that the pH value of the solution decreased with the increase of the discharge time and nitrogen deposited in the solution in the form of $NO_2^-$, $NO_3^-$ and $-NH_2$.

So far, the radiation chemistry of water has been considered. It is possible that energy of implanted ions can be deposited directly in the biological molecules of interest (e.g., DNA), but the intermediaries duo to the ionization of solvent molecules will also exert indirect influence on the organism. It is primarily the radical that is responsible for the indirect influence[29,30].

### 3. Biomolecular damage by radical

As stated above, Creating radicals from latent, stable sources is a key feature of radiation chemistry. Many biological effects induced by radiation rely on short-lived and high-reactive chemical intermediates. DNA and protein are relatively easily attacked by these intermediates, in particular free radicals (mainly OH radical, hydrogen atom, and hydrated electron) that are a constant presence in biological systems owing to the action of ionizing radiation.

### 3.1. DNA and its components damage by radicals

The long-standing interest in radiation biology at the molecular level has firstly been on DNA and its building blocks (e.g., nucleobases and nucleotides) because of their primary genetic importance and their high-sensitivity to ionizing radiation. Experiments with scavengers specific for individual reactive species have shown that it is primarily the radical, especially the hydroxyl radical (OH·), that is responsible for radiation damage to DNA[31,32]. In aqueous solution, the nucleobases react with the OH radical and the solvated electron at close to diffusion-controlled rates, while the H atom reacts one order of magnitude more slowly, and in nucleosides and nucleotides, of these three species only the OH radical shows a moderate inclination of reacting with the sugar moiety[33].

### 3.1.1. Nuleobases

Radicals are "spooky" forms of chemicals. The hydroxyl radical (OH·) is the most chemically reactive one of all reactive oxygen species (ROS), with short lifetime and high rate constant. In the case of purines and pyrimidines, OH-radical attack is by addition to the ring double bond, a reflection of its electrophilic nature. The rate constants for reaction of OH· with the electron-rich iminazole are higher compared to those with the electron-deficient pyrimidine [34], although there is debate over this opinion[35]. For pyrimidines, the OH radical prefers to attach to the $C_5$-$C_6$ double bond, mainly at the $C_5$ site of relatively high electron density[35-38].

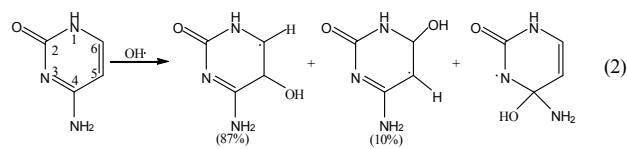
(2)

For purines, the OH radical addition occurs at two sites, i.e., $C_4$ and $C_8$ (eq. 3)[39,40]. A major part of the OH radicals add at the $C_4$-$C_5$ double bond of the purine system, and another universal occurrence is a ring-opening reaction of the adduct formed by attachment of OH· to $C_8$ (eq. 4)[41].

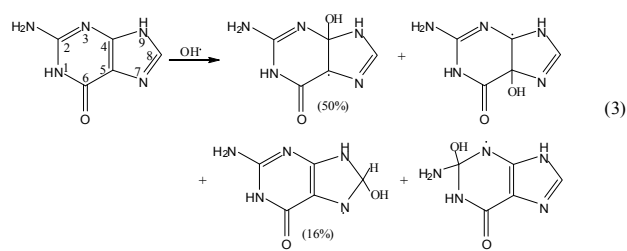
(3)

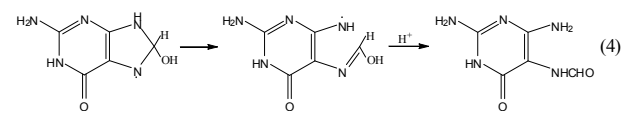
(4)

OH radical induced H-abstraction can also happen to the methyl group in, e.g., thymine, which led to the formation of allylic radical of small yield[32]. H-atom abstraction at heteroatom N, amino N, and methane C is negligible.

Similar to the OH radical, the H-atom acts as an electrophilic one, and in its addition to C=C double bonds it has a preference for the electron-richer site[33,42].

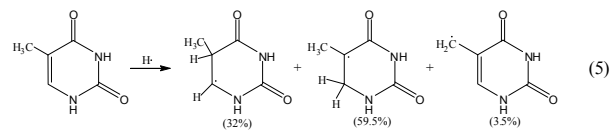
(5)

The pattern of the site preference is influenced by the species of pyrimidines in ways different from the OH radical[42,43]. The third product in eq. 5 is due to the allylic H-abstraction by the H-atom as compared to the OH radical, and this process is of small importance. Little is know about the reactions of the H atom in the purine systems. However, being an electrophilic radical also, the position-specificity for the addition of the H atom is expected to be similar to that of the OH radical in so far as the carbon atoms of the purine skeleton are considered. Dertinger[44] has put forward a mechanism for the reaction of the H atoms with the base moiety of the purine nucleotides with the result that yielded the adducts at $C_8$.

The nucleobases have a very high intrinsic reactivity with $e_{aq}^-$ endowed by the electron-deficient pyrimidine ring[39]. They capture radiation induced hydrated electrons to form short-lived radical anions, and then protonation reactions occur to their corresponding $e_{aq}^-$ adducts[33,39,42].

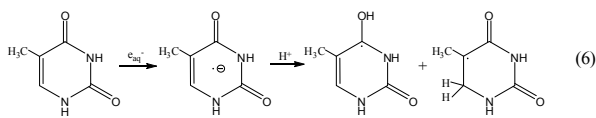 (6)

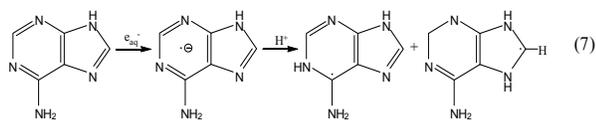 (7)

The properties of the radical anions of the purines formed by their reaction with the solvated electrons (eq. 7) are very different from those of the thymine radical anion (eq. 6), but show some resemblance to those of the cytosine radical anion. The protonation of purines is usually fastest at the position of a heteroatom (eq. 7), followed by a "rearrangement" that results in protonation at $C_2$ and $C_8$ of the purine system[39,45,46].

It has been the aim of investigators not only to learn about the reactivity of the nucleobases but also to identify their reaction products with OH·, H·, and $e_{aq}^-$. Khattak and Green[47] have studied the γ-ray irradiated aqueous solutions of cytosine, and five different constituents were separated, i.e., uracil (a deaminated hydroxylation product), 5 and 6-hydroxy cytosines, 5,6-dihydroxy cytosine and hydroxy-hydro cytosine. The irradiation of thymine in $N_2O$ saturated aqueous solution by 60MeV $^{18}O^{8+}$ ion beam give rise to a variety of different kinds of products, including hydroxyl and hydro adducts and dimmers[48]. In the presence of $O_2$, irradiation $G$-values for nucleotide destruction are increased 2–3 times[49], and most radicals are converted into the corresponding peroxyl form[50], but they will ultimately go back to their molecular form through the loss of a $H_2O_2$ molecule[51,52].

### 3.1.2. Deoxyribose

When the radiation induced radicals attack the DNA, nearly all H atoms and the major part of the OH radicals add to the double bonds of the bases but 10-20% of the OH radicals react with the sugar moiety by abstracting hydrogen[53]. The H-abstractions from $C_{1'}$, $C_{3'}$, $C_{4'}$ positions require similar energies and is a little more favorable than from $C_{2'}$.[54] After a deliberate consideration on the basis of their bond strength and their accessibility in nucleotides and in DNA, the results are in general agreement with the observation that the major product of H-abstraction in DNA is the $C_{4'}$ radical[55].

Two possible routes have been postulated for the cleavage of the phosphate ester bond and the ring-open of the ribose of the $C_{4'}$ radical[56-58]. The reactions in eq. 8 show the chemical steps of the radiation induced DNA strand break. Ullrich and Hagen[59] demonstrated that about 60 percent of the liberated DNA material consists of nucleosides, part of which carries a damaged sugar molecule.

### 3.1.3. Nucleosides and Nucleotides

The aggressive radical reactions aimed at nucleobases can also happen to their corresponding nucleosides and nucleotides in the same way[40], for example, the OH radical addition to $C_4$ and $C_8$ of the purine system. With the cytosine nucleosides and nucleotides the probability of OH attack at the cytosine molecule is >80%[38]. However, there exist rate-decreasing effects on the larger biomolecules, i.e., the lower values for reaction with the nucleosides and nucleotides compared with the free bases probably reflect the decreased electron density due to the electron-withdrawing ribose substituent at $N_9$[40,41]. The nucleosides also have a high reactivity with $e_{aq}^-$[60], but on introduction of one phosphate group into the system, the rate constants decrease[39].

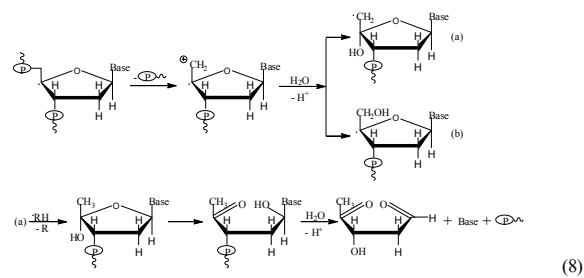

(8)

As mentioned in section 3.1.2 (eq. 8), OH radical induced H-abstractions result in the liberation of the bases and the cleavage of both 3' and 5' phosphate ester bonds. Radicals on the sugar moiety of the DNA backbone contribute to a significant extent to strand breaks. In general, strand breaks occur through cleavage of the phosphate ester bond, a process which is almost certainly initiated by one or more precursor radicals on the sugar moiety[61,62]. These reactions are equivalent to the brand-break-producing reactions in DNA.

In order to investigate the influence of the inter-nucleotide linkages on the radiation effects on mononucleotides, some polynucleotides are employed[63-69]. To a large extent the nucleobases OH adduct radicals attack the sugar moiety of polynucleotides thereby inducing base release and strand breakage[65,67,69], and Spin-trapped radicals exhibite broad ESR lines[64]. Deeble and von Sonntag proposed that the OH- and H-adduct radicals react with the sugar moiety of a different nucleotide also with the consequence of base release and strand scission [66]. By varying the experimental conditions of poly(U) radiation research, the respective efficiencies with which OH, H and $e_{aq}^-$ release uracil were found to be 51, 28 and about 3 per cent. This indicates the important role of the polymeric structure in the mechanism leading to efficient base release. Strand breakage is connected with the formation of phosphomonoester end groups, and the G-values are 0.47, 0.17, 0.04, 0.04 for poly(U), poly(C), poly(A) and poly(G), repectively[68].

### 3.1.4. DNA

Extrapolation of the effects of radiation upon nucleotides to effects upon deoxyribonucleic acid is complicated in an uncertain way by the influence of long inter-nucleotide linkages and hydrogen bonding of base pairs. Ionizing radiations are known to randomly ionize all available sites. However, the initial radicals found by ESR are the radical anions and cations of the DNA bases[68,70]. Ab initio molecular orbital calculations also verify this point of view[71]. The trend in Koopmans ionization potential among DNA building blocks can be summarized as base < deoxyribose < phosphate. It clearly appears from this trend that the four natural DNA bases will be more easily ionized than any other component of the DNA molecules.

It has been estimated that the water radicals contribute about 60% to the cellular-DNA radiation damage[9,10], with different types of lesions including base and sugar damages, strand denaturation, double strand breaks (DSBs), single strand breaks (SSBs), DNA-DNA and protein-DNA crosslinks. For ionizing radiation, studies with scavenger molecules indicate that almost all of the indirect damage to DNA is due to attack by the highly reactive hydroxyl radical (OH·)[9,72]. Experiment with the enzyme endonuclease Ⅲ, which recognizes a number of oxidized pyrimidines in DNA and converts them to strand breaks, also show the evidence for the oxidative base damage formed by the action of OH·[73]. Removal of DNA-bound proteins give an additional increase in ssbs and dsbs by a factor of 14 and 5 respectively and the increase in both ssbs and dsbs are caused by OH radicals[74]. The reducing counterparts of OH· (H· and $e_{aq}^-$) are relatively ineffective, especially at inducing DNA strand breaks[75].

As we all know, in its quest to achieve stability, the free radical tries to either pass on the extra electron to its nearest neighbour, or grab an electron from the neighbour to make up a pair to revolve around its own nucleus. This causes the neighbour to become a free radical, sometimes setting off a chain reaction. The water radicals randomly attack all available sites of DNA, but the breakpoints of DNA are non-random[76,77]. Radiation-induced electron migration along DNA is a mechanism by which randomly produced stochastic energy deposition events can lead to non-random types of damage along DNA manifested distal to the sites of the initial energy deposition[78].

Electron transfer is a common phenomenon along the chains of peptides[79,80] and DNA[39,81-83]. Electron transfer often coupled to (de)protonation. The two reactions may occur one before the other in a stepwise mechanism, or in a single, concerted step. Because of proton-coupled electron transfer (PCET), there may be relatively sensitive sites to ionizing in DNA molecules, and as a result, the nonrandom distribution of breaks is formed. The site of ultimate deposition of positive and negative charge and unpaired spin is obviously of great importance in understanding the mechanism of radiation-induced damage to DNA. Results from ESR experiments have led to the development of the two-component hypothesis that the radical cations end up with the purines, particularly with the guanine moiety[84], whereas the final site of deposition of the ejected electron is with the pyrimidines, particularly with cytosine[85,86]. The driving force for such intramolecular electron "hops" is the difference in the reduction potentials of the DNA base radicals. The guanine radical has the lowest reduction potential and the ease of oxidation follow the trend G>A>C>T[87]. Therefore, guanine is the target for ionization damage in DNA. This have been demonstrated theoretically[88] and experimentally[89]. In the improvement studies, Saito[90] and Henderson[91] indicated that the most electron-donating and readily reakable sites in duplex DNA are the guanine residues located 5' to guanine at GG steps, due to the π-stacking interaction of the two guanine bases.

### 3.2 Amino acids, peptides, and proteins damage by radicals

Although DNA is the primary genetic material and therefore a comprehensive literature on the radiation chemistry of DNA and constituent compounds has been developed, detailed knowledge of reaction mechanisms in the radiolysis of proteins is also of great importance because they play key roles in the processes of gene expression and regulation. Radicals produced in proteins by active radiolytic species of water, $e_{aq}^-$, ·OH, ·H, etc.(eq. 1), are quite conceivably the primary species responsible for radiation effects on proteins. The radical intermediates produced by ionizing radiation in aqueous solutions of many amino acids have been observed by ESR spectroscopy during in situ radiolysis[92,93] or in frozen glasses[94,95].

#### 3.2.1. Amino acids

The hydroxyl radical (OH·) can react with amines and amino acids by abstracting a carbon-bound H-atom.

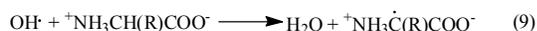

OH· + $^+NH_3CH(R)COO^-$ ⟶ $H_2O$ + $^+NH_3\dot{C}(R)COO^-$     (9)

ESR (Electronic Spin Resonance) has been the most common analytic technique since the pioneering work of Morton from the late 1950's[96]. The radicals produced by reaction of hydroxyl radicals with some amines, amino acids, and related compounds in aqueous solutions have been studied by ESR[97]. Rustgi and his coworkers[98] have investigated the radicals by reaction of hydroxyl radicals with nineteen amino acids in aqueous solution, and they found only side-chain radicals were identified for alanine, threonine, aspartic acid, asparagine, lysine, phenylalanine, tyrosine, proline and hydroxyproline; whereas for glycine the C(2) carbon radical was spin-trapped. Both C(2) carbon radicals and side-chain radicals were assigned to valine, leucine, isoleucine, serine, glutamic acid, glutamine, arginine and methionine. For acyl amides(e.g., acetamide, succinamide), the observed radicals are formed by H-abstraction from the carbon atoms attached to the carbonyl group, and for their N-alkylated derivatives, H-abstraction occurs mainly from the N-methyl group[99,100]. The identification of the site of OH· attack on N-methyl amides is helpful for the study of radical formation in peptides and proteins.

The reactions by which short-lived radicals are produced in irradiated aqueous solutions of the amino acids have been found to be deamination by hydrated electrons as well as H-abstraction by hydroxyl radicals[95,101,102]. For amino acids with reactive side groups (e.g., asparagine, methionine, and cystine), the reaction of hydrated electrons shows that the reactive side group in these molecules is a point of electron localization of attack; for amino acids with alkyl side groups, the electron is found to add to the carboxyl group, the amino acids dianion formed subsequently deaminates to produce a $\dot{C}HRCO_2^-$ radical (eq. 10),

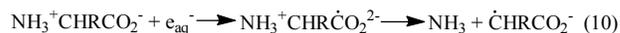

$NH_3^+CHRCO_2^- + e_{aq}^- \longrightarrow NH_3^+CHR\dot{C}O_2^{2-} \longrightarrow NH_3 + \dot{C}HRCO_2^-$     (10)

and this species abstracts H-atom from the α carbon of another amino acid to form the $RCH_2CO_2^-$ molecule[94]. Rustgi and his colleagues[103] identified radicals produced by reductive deamination of 19 amino acids. The scission of the $CH_3-S-$ and -$S-CH_2-$ bonds of methionine and the opening of the cyclic structure of proline were also observed in their experiment. A hydroxyl group in the β-position to the amino group promotes deamination of amino alcohols, amino acids, di- and tripeptides[104]. The radical-induced damage of two simplest amino acids, i.e., glycine and alanine, have been the subjects of much attention, theoretically[105-107] and experimentally[108,109]. The attention is also focused on the cysteine molecule[93] because it can form disulfide bond which plays a crucial role in the formation of protein secondary structure and the protein folding. The general results are shown in Table 2.

**Table 2.** Radical reaction scheme in aqueous glycine and cystine solutions.

| Glycine ($NH_3^+CH_2COO^-$) | Cysteine (CySH) |
|---|---|
| (1)$e_{aq}^-$+$NH_3^+CH_2COO^-$→$NH_3$+·$CH_2COO^-$ | (1)$e_{aq}^-$+CySH→Cy·+SH· |
| (2)H·+ $NH_3^+CH_2COO^-$→$H_2$+ $NH_3^+\dot{C}HCOO^-$ | (2)H$\dot{C}$+CySH→$H_2$+CyS· |
| (3)OH·+ $NH_3^+CH_2COO^-$→$H_2O$+ $NH_3^+\dot{C}HCOO^-$ | (3)H·+CySH→$H_2S$+Cy· |
| (4)$H_2O_2$+ $NH_3^+\dot{C}HCOO^-$→$NH_2$=CHCOO·+$H_2O$+OH | (4)OH·+CySH→$H_2O$+CyS· |
| (5)2 $NH_3^+\dot{C}HCOO^-$→$NH_2^+$=CHCOO·+ $NH_3^+CH_2COO^-$ | (5)$O_2^-$+CySH→$HO_2^-$+CyS· |
| (6)$NH_2^+$=CHCOO·+$H_2O$→$NH_4^+$+OHCCOO· | (6)Cy·+CySH→CyH+CyS· |
| (7)$NH_2^+$=CHCOO·+$H_2O$→$NH_3$+HCHO+$CO_2$ | (7)2CyS·→CySSCy |
| (8)$NH_3^+\dot{C}HCOO^-$+·$CH_2COO^-$→ $NH_2^+$=CHCOO·+$CH_3COO^-$ | (8)$H_2O_2$+2CySH→ $2H_2O$+CySSCy |

#### 3.2.2. Peptides

H-abstraction and N-terminal deamination definitely also occur in peptides[110-113]. In reactions of hydroxyl radicals with peptides, both backbone and side-chain radicals were detected[112,113]. Neta[92] and Sevilla[110] reported that protonation of the amine group will lower the activation energy for deamination in amino acid and peptide anions.

In addition to the above two well-known radiation effects, C-N bond scission was also observed at three sites[114]. Besides the peptide bond breakage, cleavage occurs between the nitrogen of the ammonium group and the α-carbon and between the nitrogen of the peptide linkage and the adjoining α-carbons (eq. 11)[113,115].

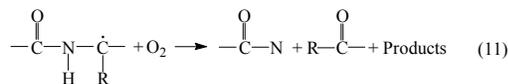

The presence of hydroxyl groups in side residues promotes destruction of the peptide chain with formation of amides of amino acids, while suppression of deamination and main chain destruction processes in these compounds is observed in the presence of oxygen[104]. It is reported that OH radicals can be added to the aromatic cycles of the aromatic unsaturated residues (e.g., tryptophan and histidine)[113,116]. For the tryptophan residue, the radiolytic oxidation arises predominantly through reaction initiated by OH addition to unsaturated bonds of the indole moiety which leads to formation of formylkynurenine as a major degradation product (eq. 12)[113,117].

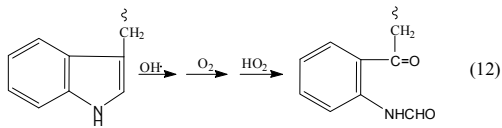

This may be the explanation for the marked protection of the aliphatic chain by the aromatic ring system reported by Casteleijn[118].

### 3.2.3. Proteins

The reactions of OH· at α carbon along the protein main chain were found to lead to oxidative degradation to yield amide and keto acid functions (eq. 13, refer to eqs. 9 and 11)[113,119]. The separation and isolation of protein fragments formed by main chain cleavage has been achieved in the radiolysis of lactate dehydrogenase (LDH)[120] and bovine serum albumen (BSA)[121] in oxygenated solution. Whereas in oxygen-free solutions, the carbonyl group of the peptide bond represents a major trapping center for $e_{aq}^-$[111, 114], which also lead to the main chain scission in protein (eq. 14)[122].

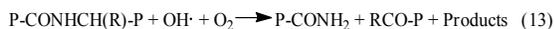
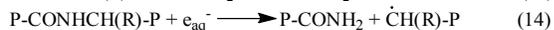

$$\text{P-CONHCH(R)-P} + \text{OH·} + O_2 \longrightarrow \text{P-CONH}_2 + \text{RCO-P} + \text{Products} \quad (13)$$
$$\text{P-CONHCH(R)-P} + e_{aq}^- \longrightarrow \text{P-CONH}_2 + \dot{\text{C}}\text{H(R)-P} \quad (14)$$

For proteins with sulfur-containing side chain, such as urease[123], Papain[124], and glyceraldehyde-3-phosphate dehydrogenase[125], three reaction models for the radiation-induced oxidation of protein SH groups in oxygenated solutions have been formulated, i.e.

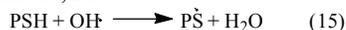
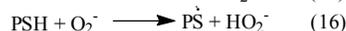
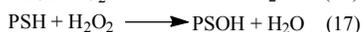

$$\text{PSH} + \text{OH·} \longrightarrow \text{P}\dot{\text{S}} + H_2O \quad (15)$$
$$\text{PSH} + O_2^- \longrightarrow \text{P}\dot{\text{S}} + HO_2^- \quad (16)$$
$$\text{PSH} + H_2O_2 \longrightarrow \text{PSOH} + H_2O \quad (17)$$

Subsequent reaction of macromolecular PS· with $O_2$ leads to formation of nonrepairable products which are presumed to be sulfinic and sulfonic derivatives. Within proteins, the disulfide bond is more sensitive to electron attack than the peptide bond. Electron trapping at disulfide bridges in proteins yield the short-lived adduct (P$\overline{SS}$P) which will undergo dissociation[113,126,127].

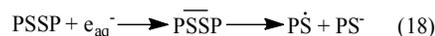

$$\text{PSSP} + e_{aq}^- \longrightarrow \text{P}\overline{\text{SS}}\text{P} \longrightarrow \text{P}\dot{\text{S}} + \text{PS}^- \quad (18)$$

The reverse reaction can also happen to form disulfide radical anion. In the presence of $O_2$, thiyl radicals readily dimerise, and thereby give rise to inter- or intramolecular protein crosslink[9].

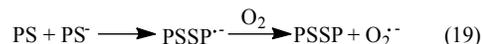

$$\text{PS} + \text{PS}^- \longrightarrow \text{PSSP}^{·-} \xrightarrow{O_2} \text{PSSP} + O_2^{·-} \quad (19)$$

There is growing evidence that proteins are early targets of reactive oxygen species[128]. Exposure of proteins to radical attack, especially in the presence of $O_2$, results in multiple changes in the target molecules, including oxidation of side-chain, backbone fragmentation, cross-linking, unfolding, changes in hydrophobicity and conformation, altered susceptibility to proteolytic enzymes, and formation of new reactive groups[129-131]. These changes are likely to initiate further potentially damaging reactions and to beget some downstream events.

## 4. The role of secondary particles

As stated in our last paper[11], the capture of electrons from the target is a significant mechanism that ionizes the target but does not produce secondary electrons. If the projectile is appropriate (e.g., He$^+$), however, a collision can eject electrons from either the projectile or the target[132]. After the charge transfer process projectile becomes either stripped or ionises a water molecule. In a stripped case, an electron is ejected in a forward direction with nearly the same velocity as the projectile. In the latter case, an ionized water molecule and an ejected electron result.

It has been reported that important damage is not due to the primary radiation itself, but due to secondary particles generated along the track after the interaction of the ionizing radiation with the biological medium. These particles may be either low-energy electrons, radicals, or singly and multiply charged ions. In that sense, a lot of experiments have been performed recently in order to investigate the action of these secondary particles on biologically relevant molecules. When a charged particle crosses the cell, secondary particles such as low energy electrons, radicals and (singly/multiply charged) ions are formed within the track. The interaction of these secondary particles with biologically relevant molecules is responsible for a large fraction of the induced biological damage in the cell.

Some collision models of secondary ions and biomolecules have been introduced in our last contribution[11]. However, low energy electrons (0–20 eV) are the most abundant secondary particles [20]. Recently, extensive research has been devoted to their interaction with DNA and its constituents[133-151], as well as amino acid[152]. In particular, a great number of studies involving the interaction with low-energy electrons have been undertaken on supercoiled DNA, but also on DNA and RNA bases. Low energy electrons were found to cause single and double strand breaks of plasmid DNA [133-135] and it is now known that even electrons of almost zero kinetic energy can induce single strand breaks [136]. Interestingly, very similar resonances as found for DNA single and double strand breaks as a function of electron energy have been observed in electron-induced fragmentation of nucleobases[137–151].

Ionization and fragmentation of DNA and its constituents is a primary step in biological radiation damage. The interaction of these secondary particles with biologically relevant molecules is responsible for a large fraction of the induced biological damage in the cell.

**References**

[1] Yu Zengliang(*Ed.*), Yu Liangdeng, T. Vilaithong, and I. Brown (*Trans.*), Introduction to Ion Beam Biotechnology, Springer Science+Business Media, New York, 2006, p.9.
[2] Yu Z.L., Wu Y.J., Deng J.G. *et al.*, Mutation breeding by ion implantation. *Nucl. Instru .Meth. Phys. Res. B.,* 1991, 59/60:705-708
[3] Yu Z.L., Yang, J.B., Wu, Y.J. *et al.* Transferring Gus gene into intact cells by low energy ion beam. *Nucl. Instru. Meth. Phys. Res. B.,* 1993, 80/81:1328-1331.
[4] Yu, Z.L., Ion beam application in genetic modification, IEEE Trans. Plas. Sci., 2000,28(1):128-132.
[5] L.F. Wu, Z.L. Yu, Radiobiological effects of a low-energy ion beam on wheat, Radiat. Environ. Biophys., 2001, 40(1):53-57.
[6] B. Rydberg, DNA strand breaks induced by low-energy heavy ions, Inter. J. Radiat. Biol. 1985,47(1):57-61.
[7] Y. Chen, B.Y. Jiang, Y.S. Chen, *et al.*, Formation of plasmid DNA strand breaks induced by low-energy ion beam: indication of nuclear stopping effects, Radiat. Environ. Biophys., 1998,37(2):101-106.
[8] C.X. Xie, J.H. Guo, Z.L. Yu, *et al.*, Evidence for base substitutions and repair of DNA mismatch damage induced by low energy $N^+$ beam implantation in *E. Coli*, High Tech. Lett., 2003,9(2):1-6.
[9] C. von Sonntag, The Chemical Basis of Radiation Biology, Taylor and Francis, London, 1987.
[10] D. Michael, P. O'Neill, Molecular biology: a sting in the tail of electron tracks. Science. 2000,287:1603-1604.
[11] http://cn.arxiv.org/abs/0807.0079.
[12] Ferradini, C., Jay-Gerin, J.-P., 1999. Radiolysis of water and aqueous solutions-History and recent state of the science. Can. J. Chem.,77:1542-1575.
[13] LaVerne J.A., Development of radiation chemistry studies of aqueous solutions with heavy ions. Nucl. Instru. Meth. B. 1996, 107:302, and references therein.
[14] W.M. Zhang, Introduction to radiation chemistry. USTC Press, Hefei, 1993.
[15] A. Appleby and Harold A. Schwarz, Radical and molecular yields in water irradiated by γ rays and heavy ions, J. Phys. Chem., 1969, 73(6):1937-1941.
[16] LaVerne, J. A. (1989) The production of OH radicals in the radiolysis of water with $^4$He ions. *Radiat. Res.*, **118**, 201–210.
[17] Jay A. Laverne and Hiroko Yoshida, Production of the Hydrated Electron in the Radiolysis of Water with Helium Ions, *J. Phys. Chem. 1993,97,* 10720-10724.
[18] Baldacchino G, Le Parc D, Hickel B, Gardes-Albert M, Direct observation of $HO_2/O_2^-$ free radicals generated in water by a high-linear energy transfer pulsed heavy-ion beam. Radiat Res. 1998 ,149(2):128-33.
[19] Frongillo Y, Fraser M J, Cobut V, Goulet T, Jay-Gerin J P and Patau J P 1996 Evolution of the species produced by slowing down of fast protons in liquid water: simulation based on the independent reaction times approximation *J. Chim. Phys.* **93** 93–102.
[20] V. Cobut, Y. Frongillo, J. P. Patau, T. Goulet, M. -J. Fraser and J. -P. Jay-Gerin, Monte Carlo simulation of fast electron and proton tracks in liquid water -- I. physical and physicochemical aspects , *Radiat. Phys. Chem.,* 51( 3), 1998:229-243.
[21] Y. Frongillo, T. Goulet, M.-J. Fraser, V. Cobut, J.P. Patau, Monte Carlo simulation of fast electron and proton tracks in liquid water-II .nonhomogeneous chemistry. Radiat. Phys. Chem. 1998,51(3):245-254.
[22] B. Gervais, M. Beuve, G.H. Olivera, M.E. Galassi, Numerical simulation of multiple ionization and high LET effects in liquid water radiolysis. Radiation Physics and Chemistry 75 (2006) 493–513.
[23] N.J.B. Green, M.J. Pilling, S.M. Pimblot, P. Clifford, Stochastic Modeling of Fast kinetics in a radiation track. J. Phy. Chem. 1990, 94:251-258.
[24] Uehara S., Nikjoo H., Monte Carlo simulation of water radiolysis for low-energy charged particles. J. Radiat. Res. 2006, 47(1):69-81.
[25] Turner J. E., Atoms, Radiation, and Radiation Protection, 2nd ed. New York, Wiley-Intersceince, 1995.
[26] H. Kudoh and Y. Katsumura. 2001. Ion-beam radiation chemistry. In: C.D. Jonah, B.S.M. Rao, editors. Radiation Chemistry: Present Status and Future Trends. Elsevier Sciences B.V. pp37-66.
[27] X.Q. Wang, J.W. Han, W. Wang, Synthesis of bio-organic molecules of life induced by low energy ion implantation. Chin. Phys. Lett. 2004, 21(10):1941-1943.
[28] C.L. Shao, J.M. Yao, X.Q. Wang, $N^+$ ions irradiation-induced mass deposition in sodium carboxylic molecules, Radiat. Phys. Chem., 1998, 51(2):117-120.
[29] R.F. Cook, The effects of water and a protective agent on gamma-ray induced free radicals in mustard seeds. Inter. J. Radiat. Biol. 1963, 7(5):497-504.
[30] Deng J.G, Yu Z.L., ESR study on seeds irradiated with nitrogen ions. Nucl. Tech., 1992, 15(4):245-250 (in Chinese).
[31] Siddiqi M.A., Bothe E., Single- and double-strand break formation in DNA irradiated in aqueous solution: dependence on dose and OH radical scavenger concentration. Radiat. Res. 1987, 112(3): 449-463.
[32] Milligan J.R., Aguilera J.A., Ward J.F., Variation of single-strand break yield with scavenger concentration for plasmid DNA irradiated in aqueous solution. Radiat. Res. 1993,133(2):151-157.
[33] C. von Sonntag and H.-P. Schuchmann. 2001. Radiation Chemistry of the Nucleobases. In: C.D. Jonah, B.S.M. Rao, editors. Radiation Chemistry: Present Status and Future Trends. Elsevier Sciences B.V. pp513-552.
[34] G.V. Buxton, C.L. Greenstock, W.P. Helman, A.B. Ross, Critical Review of Rate Constants for Reactions of Hydrated Electrons, Hydrogen Atoms and Hydroxyl Radicals in Aqueous Solution. J. Phys. Chem. Ref. Data. 1988,17:513-886.
[35] M.G. Ormerod, B.B. Singh, Reaction of OH Radicals with Nucleic Acid Bases-An E.S.R. Study. Inter. J. Radiat. Biol. 1966, 10(6):533-538
[36] J. Planinić, E.S.R. Studies of Free Radicals Derived from Reaction of OH with Uracil and Thymine. 1980,38(6):651-659.
[37] S. Fujita, S. Steenken, Pattern of OH radical addition to uracil and methyl- and carboxyl-substituted uracils. Electron transfer of OH adducts with *N,N,N',N'*-tetrsmethyl-*p*-phenylenediamine and tetranitromethane. J. Am. Chem. Soc. 1981,103:2540-2545.
[38] D.K. Hazra, S. Steenken, Pattern of OH radical addition to cytosine and 1-, 3-, 5-, and 6-substituted cytosines. Electron transfer and dehydration reactions of the OH adducts. J. Am. Chem. Soc. 1983,105:4380-4386.
[39] S. Steenken, Purine bases, nucleosides, and nucleotides: aqueous solution redox chemistry and transformation reactions of their radical cations and $e^-$ and OH adducts. Chem. Rev. 1989,89:503-520.
[40] A.J.S.C. Vieira, S. Steenken, Pattern of OH radical reaction with adenine and its nucleosides and nucleotides. Characterization of two types of isomeric OH adducts and their unimolecular transformation reactions. J. Am. Chem. Soc. 1990,112:6986-6994.
[41] A.J.S.C. Vieira, S. Steenken, Pattern of OH radical reaction with 6- and 9-substituted purines. Effect of substituents on the rates and activation parameters of the unimolecular transformation reactions of two isomeric OH adducts. J. Phys. Chem. 1987,91:4138-4144.
[42] D.J. Deeble, S. Das, C. von Sonntag, Uracil derivatives: sites and kinetics of protonation of the radical anions and the UV spectra of the C(5) and C(6) H-atom adducts. J. Phys. Chem. 1985,89:5784-5788.
[43] Das S, Deeble DJ, von Sonntag C, Site of H atom attack on uracil and its derivatives in aqueous solution. Z. Naturforsch. 1985,40(3-4):292-294
[44] H. Dertinger, G. Hartig, Mechanisms of radical formation in irradiated purine bases and derivatives: An E.P.R. study using computer technique. Inter. J. Radiat. Biol. 1972, 21(3):279-292.
[45] L.P. Candeias, S. Steenken, Electron adducts of adenine nucleosides and nucleotides in aqueous solution: protonation at two carbon sites ($C_2$ and $C_8$) and intra- and intermolecular catalysis by phosphate. J. Phys. Chem. 1992,96:937-944.
[46] L.P. Candeias, P. Wolf, P. O'Neill, S. Steenken, Reaction of hydrated electrons with guanine nucleosides: fast protonation on carbon of the electron adduct. J. Phys. Chem. 1992,96:10302-10307.
[47] M.N. Khattak, J.H. Green, Gamma-irradiation of nucleic-acid constituents in de-aerated aqueous solutions I. cytosine. Inter. J. Radiat. Biol. 1966, 11(2):131-136.
[48] X.C. Cai, Z.Q. Wei, W.J. Li, The products analysis of $^{18}O^{8+}$ ion beam on thymine in $N_2O$ saturated aqueous solution. Chin. J. Anal. Chem. 1999,27(8):869-874 (in Chinese).
[49] G. Hems, Effects of ionizing radiation on aqueous solutions of inosine and adenosine. Radiat. Res. 1960,13:777-787.
[50] G. Scholes, J. Weiss, Organic hydroxy-hydroperoxides: a class of hydroperoxides formed under influence of ionizing radiations. Nature, 1960,185:305-306.
[51] C. von Sonntag and H.-P. Schuchmann, Angew. Chem. Int. Ed. Engl., 30(1991) 1229.
[52] H. Loman, M. Ebert, The radiation chemistry of thymine in aqueous solution some reactions of the thymine-electron adduct. Inter. J. Radiat. Biol. 1970,18(4):369-379.
[53] Scholes G, Ward JF, Weiss J., Mechanism of the radiation-induced degradation of nucleic acids. J Mol Biol. 1960,2:379-91.
[54] K. Miaskiewicz, R. Osman, Theoretical study on the deoxyribose radicals formed by hydrogen abstraction. J. Am. Chem. Soc. 1994, 116:232-238.
[55] M. Kuwabara, Z.Y. Zhang, G. Yoshii, E.S.R. of spin-trapped radicals in aqueous solutions of pyrimidine nucleosides and nucleotides. reactions of the hydroxyl radical. Inter. J. Radiat. Biol. 1982,41(3):241-259.
[56] S. Steenken, G. Behrens, D. Schulte-Frohlinde, Radiation chemistry of DNA model compounds. Part IV. phosphate ester cleavage in radicals derived from glycerol phosphates. Inter. J. Radiat. Biol. 1974,25(2):205-210.
[57] L. Stelter, C. von Sonntag, D. Schulte-Frohlinde, Radiation chemistry of DNA-model compounds V. Pphosphate elimination from ribose-5-phosphate after OH radical attack at C-4. Inter. J. Radiat. Biol. 1974,25(5):515-519.
[58] M. Dizdaroglu, C. von Sonntag, D. Schulte-Frohlinde, Strand breaks and sugar release by γ-irradiation of DNA in aqueous solution. J. Am. Chem. Soc. 1975, 97:2277-2278.
[59] M. Ullrich, U. Hagen, Base liberation and concomitant reaction in irradiated DNA solutions. Inter. J. Radiat. Biol. 1971,19(6):507-517.
[60] K.J. Visscher, M.P. De Haas, H. Loman, Fast protonation of adenosine and of its radical anion formed by hydrated electron attack: A nanosecond optical and DC-conductivity pulse radiolysis study. Inter. J. Radiat. Biol. 1987,52(5):745-753.
[61] S. Fujita, Radiolysis of Nucleosides in Aqueous Solutions: Base Liberation by the Base Attack Mechanism. Inter. J. Radiat. Biol. 1984,45(4):371-377.
[62] J.F. Ward, I. Kuo, Deoxynucleotides-models for Studying Mechanisms of Strand Breakage in DNA II. Thymidine 3'5'-diphosphate. Inter. J. Radiat. Biol. 1973,23(6):543-557.
[63] J.F. Ward, M.M. Urist, γ-irradiation of Aqueous Solutions of Polynucleotides. Inter. J. Radiat. Biol. 1967,12(3):209-218.

[64] A. Joshi, S. Rustgi, P. Riesz, E.S.R. of Spin-trapped Radicals in Gamma-irradiated Aqueous Solutions of Nucleic Acids and Their Constituents. Inter. J. Radiat. Biol. 1976,30(2):151-170.
[65] D.G.E. Lemaire, E. Bothe, D. Schulte-Frohlinde, Yields of Radiation-induced Main Chain Scission of Poly U in Aqueous Solution: Strand Break Formation Via Base Radicals. Inter. J. Radiat. Biol. 1984,45(4):351-358.
[66] Deeble DJ, von Sonntag C., Gamma-radiolysis of poly(U) in aqueous solution. The role of primary sugar and base radicals in the release of undamaged uracil. Inter. J. Radiat. Biol. 1984,46(3):247-270.
[67] Deeble DJ, Schulz D, von Sonntag C., Reactions of OH radicals with poly(U) in deoxygenated solutions: sites of OH radical attack and the kinetics of base release. Inter. J. Radiat. Biol. 1986,49(6):915-926.
[68] Murthy CP, Deeble DJ, von Sonntag C., The formation of phosphate end groups in the radiolysis of polynucleotides in aqueous solution. Z. Naturforsch. 1988,43(7-8):572-576.
[69] C.R. Paul A1, E.E. Budzinski A1, A. Maccubbin, Characterization of Radiation-induced Damage in d(TpApCpG). Inter. J. Radiat. Biol. 1990,58(5):759-768.
[70] Gregoli S, Olast M, Bertinchamps A, Radiolytic pathways in gamma-irradiated DNA: influence of chemical and conformational factors. Radiat, Res. 1982,89(2):238-254.
[71] A. Colson, B. Besler, M.D. Sevilla, Ab initio molecular orbital calculations on DNA radical ions. 3. Ionization potential and ionization sites in components of the DNA sugar phosphate backbone. J. Phys. Chem. 1993,97:8092-8097.
[72] C.M. de Lara, T.J. Jenner, K.M. Townsend, The effect of dimethyl sulfoxide on the induction of DNA double-strand breaks in V79-4 mammalian cells by alpha particles. Radiat Res. 1995,144(1):43-49.
[73] K.M. Prise, C.H.L. Pullar, B.D. Michael, A study of endonuclease III-sensitive sites in irradiated DNA: detection of α-particle-induced oxidative damage. Carcinogenesis, 1999,20(5):905-909.
[74] J. Nygren, M. Ljungman, G. Ahnström, Chromatin Structure and Radiation-induced DNA Strand Breaks in Human Cells: Soluble Scavengers and DNA-bound Proteins Offer a Better Protection Against Single- than Double-strand Breaks. Inter. J. Radiat. Biol. 1995,68(1):11-18.
[75] R.S. Feldberg, J.A. Carew, Water Radiolysis Products and Nucleotide Damage in gama-irradiated DNA. Inter. J. Radiat. Biol. 1981,40(1):11-17
[76] M. Lobrich, P.K. Cooper, B. Rydberg, Non-random distribution of DNA double-strand breaks induced by particle irradiation. Inter. J. Radiat. Biol. 1996,70(5):493-503.
[77] H.C. Newman, K.M. Prise, M. Folkard, B.D. Michael, DNA double-strand break distributions in X-ray and alpha-particle irradiated V79 cells: evidence for non-random breakage. Inter. J. Radiat. Biol. 1997,71(4):347-363.
[78] A.F. Fuciarelli, E.C. Sisk, J.H. Miller, J.D. Zimbrick, Radiation-induced Electron Migration in Nucleic Acids, Inter. J. Radiat. Biol. 1994,66(5):505-509.
[79] K. Bobrowski, K.L. Wierzchowski, J. Holcman, M. Ciurak, Intramolecular Electron Transfer in Peptides Containing Methionine, Tryptophan and Tyrosine: A Pulse Radiolysis Study. Inter. J. Radiat. Biol. 1990,57(5):919-932.
[80] M. Sjödin, S. Styring, H. Wolpher, Switching the redox mechanism: models for proton-coupled electron transfer from tyrosine and tryptophan. J. Am. Chem. Soc. 2005,127:3855-3863.
[81] D. Dee, M.E. Baur, Charge and excitation migration in DNA. J. Chem. Phys. 1974,60(2):541-560.
[82] C.J. Murphy, M.R. Arkin, Y. Jenkins, Long-range photoinduced electron transfer through a DNA helix. Science. 1993,262:1025-1029.
[83] C.Z. Wan, T. Fiebig, S.O. Kelley, Femtosecond dynamics of DNA-mediated electron transfer. Proc. Natl. Acad. Sci. USA. 1999,96:6014-6019.
[84] L.P. Candeias, S. Steenken, Electron transfer in di(deoxy)nucleoside phosphates in aqueous solution: rapid migration of oxidative damage (via adenine) to guanine. J. Am. Chem. Soc. 1993,115:2437-2440.
[85] W.A. Bernhard, Sites of electron trapping in DNA as determined by ESR of one-electron-reduced oligonucleotides. J. Phys. Chem. 1989,93:2187-2189.
[86] S. Steenken, J.P. Telo, H.M. Novais, L.P. Candeias, One-electron-reduction potentials of pyrimidine bases, nucleosides, and nucleotides in aqueous solution. Consequences for DNA redox chemistry. J. Am. Chem. Soc. 1992,114:4701-4709.
[87] V. M. Orlov, A. N. Smirnov, Y. M. Varshavsky, Ionization potentials and electron-donor ability of nucleic acid babes and their analogues. Tetra. Lett. 1976,17(48):4377-4378.
[88] S. Steenken, S.V. Jovanovic, How easily oxidizable is DNA? One electron reduction potentials of adenosine and guanine radicals in aqueous solution. J. Am. Chem. Soc. 1997,119:617-618.
[89] T. Melvin, S.M.T. Cunniffe, P. O'Neil, Guanine is the target for direct ionization damage in DNA, as detected using excision enzymes. Nucl. Acids. Res. 1998,26(21):4935-4942.
[90] I. Saito, M. Takayama, H. Sugiyama, K. Nakatani, Photoinduced DNA cleavage via electron transfer: demonstration that guanine residues located 5' to guanine are the most electron-donating sites. J. Am. Chem. Soc. 1995,117:6406-6407.
[91] P.T. Henderson, D. Jones, G. Hampikian, Long-distance charge transport in duplex DNA: the phonon-assisted polaron-like hopping mechanism. Proc. Natl. Acad. Sci. USA. 1999,96:8353-8358.
[92] Neta P., Fessenden, R.W., Electron spin resonance study of deamination of amino acids by hydrated electrons. J. Phys. Chem. 1970, 74(11):2263-2266.
[93] Neta P., Fessenden R.W., Electron spin resonance study of radicals produced in irradiated aqueous solutions of thiols. J. Phys. Chem. 1971, 75(15):2277-2283.
[94] Sevilla M.D., Radicals formed by the reaction of electrons with amino acids in an alkaline glass. J. Phys. Chem. 1970, 74(10):2096-2102.
[95] Sevilla M.D., D'Arcy J.B., Suryanarayana D., Conformational effects on the electron spin resonance spectra of radicals produced by electron attachment to amino acids and peptides. J. Phys. Chem. 1978, 82(24): 2589-2574.
[96] J.R. Morton, Electron spin resonance spectra of oriented radicals. Chem. Rev. 1964,64:453-471.
[97] Neta P., Fessenden R.W., Electron spin resonance study of radicals produced in irradiated aqueous solutions of amines and amino acids. J. Phys. Chem. 1971, 75(6):738-748.
[98] S. Rustgi, A. Joshi, H. Moss, P. Riesz, E.S.R. of spin-trapped radicals in aqueous solution of amino acids: Reactions of the hydroxyl radical. Inter. J. Radiat. Biol. 1977, 31(5):415-440.
[99] S. Rustgi, P. Riesz, E.S.R. Study of spin-trapped radicals formed during the photolysis of aqueous solutions of acid amides and H2O2, Inter. J. Radiat. Biol. 1978, 33(4):325-339.
[100] S. Rustgi, P. Riesz, Free Radicals in U.V.-irradiated aqueous solutions of substituted amides: an E.S.R. and spin-trapping study, Inter. J. Radiat. Biol. 1978, 34(2):149-163.
[101] P. Neta, M. Simic, E. Hayon, Pulse radiolysis of aliphatic acids in aqueous solution. III.simple amino acids. J. Phys. Chem. 1970, 74(6):1214-1220.
[102] K. Makino, F. Moriya, H. Hatano, Application of the spin-trap HPLC-ESR method to radiation chemistry of amino acids in aqueous solutions. Radiat, Phys. Chem., 1984, 23(1-2):217-228.
[103] S. Rustgi, A. Joshi, P. Riesz, F. Friedberg, E.S.R. of Spin-trapped radicals in aqueous solution of amino acids. Reactions of the hydrated electron. Inter. J. Radiat. Biol. 1977, 32(6):533-552.
[104] O. I. Shadyro, A. A. Sosnovskaya, O. N. Vrublevskaya, C-N bond cleavage reactions on the radiolysis of amino-containing organic compounds and their derivatives in aqueous solutions. Inter. J. Radiat. Biol. 2003, 79(4):269-279.
[105] P. Lahorte, F. De Proft, G. Vanhaelewyn, Density functional calculations of hyperfine coupling constants in alanine-derived radicals. J. Phys. Chem. A 1999, 103, 6650-6657.
[106] F. Ban, S. D. Wetmore, R. J. Boyd, A density-functional theory investigation of the radiation products of L-α-alanine. J. Phys. Chem. A 1999, 103, 4303-4308.
[107] F. Ban, J. W. Gauld, R. J. Boyd, A density functional theory study of the radiation products of glycine. J. Phys. Chem. A 2000, 104, 5080-5086
[108] F. Moriya, K. Maklno, N. Suzukl, Studies on spin-trapped radicals in γ-Irradiated aqueous solutlons of glycine and L-alanine by high-performance liquid chromatography and ESR spectroscopy. J. Phys. Chem. 1980, 84, 3085-3090.
[109] G. L. Hug, R. W. Fessenden, Identification of radicals and determination of their yields in the radiolytic oxidation of glycine: Time-resolved ESR methodology. J. Phys. Chem. A 2000, 104, 7021-7029.
[110] M. D. Sevilla, V.L.Brooks, Radicals formed by the reaction of electrons with amino acids and peptides in a neutral aqueous glass. J. Phys. Chem.1973,77(25):2954-2959.
[111] S. Rustgi, P. Riesz, E.S.R. and spin-trapping studies of the reactions of hydrated electrons with dipeptides, Inter. J. Radiat. Biol. 1978,34(2):127-148.
[112] A. Joshi, S. Rustgi, H. Moss, P. Riesz, E.S.R. of spin-trapped radicals in aqueous solutions of peptides: reactions of the hydroxyl radical. Inter. J. Radiat. Biol. 1978,33(3):205-229.
[113] W. M. Garrison, Reaction mechanisms in the radiolysis of peptides, polypeptides, and proteins. Chem. Rev. 1987. 87:381-398.
[114] S. Rustgi, P. Riesz, Hydrated electron-initiated main-chain scission in peptides: An E.S.R. and spin-trapping study. Inter. J. Radiat. Biol. 1978,34(5):449-460.
[115] P. Riesz, S. Rustgi, Aqueous radiation chemistry of protein and nucleic acid constituents: ESR and spin-trapping studies. Radiat. Phys. Chem.,1979,13(1-2):21-40.
[116] I. Balakrishnan, M. P. Reddy, Mechanism of reaction of hydroxyl radicals with benzene in the γ radiolysis of the aerated aqueous benzene system. J. Phys. Chem. 1970,74(4):850-855.
[117] R.V. Winchester, K.R. Lynn, X- and γ-radiolysis of some tryptophan dipeptides. Inter. J. Radiat. Biol. 1970,17(6):541-548.
[118] G. Casteleijn, J. Depireux, A. Müller, Electron-spin-resonance study of radiation-induced radicals in aromatic and aliphatic amino acids and peptides. Inter. J. Radiat. Biol. 1964,8(2):157-164.
[119] W. M. Garrison, M. Kland-English, H. A. Sokol, M. E. Jayko, Radiolytic degradation of the peptide main chain in dilute aqueous solution containing oxygen. J. Phys. Chem, 1970,74 (26):4506-4509.
[120] H. Schuessler, A. Herget, Oxygen effect in the radiolysis of proteins I. lactate dehydrogenase. Inter. J. Radiat. Biol. 1980,37(1):71-80.
[121] H. Schuessler, A. Herget, Oxygen effect in the radiolysis of proteins: Part 2 bovine serum albumin. Inter. J. Radiat. Biol. 1984,45(3):267-281.
[122] O. Inanami, M. Kuwabara, M. Hayashi, G. Yoshii, Reaction of the hydrated electron with histone H1 and related compounds studied by e.s.r. and spin-trapping. Inter. J. Radiat. Biol. 1986,49(1):47-56.
[123] S.E. Lewis, E.D. Wills, A. Wormall, Inactivation of urease by X-rays. Inter. J. Radiat. Biol. 1961,3(6):647-656.
[124] W.S. Lin, J.R. Clement, G.M. Gaucher, D.A. Armstrong, Repairable and nonrepairable inactivation of irradiated aqueous papain: effects of OH, $O_2^-$, $e_{aq}^-$, and $H_2O_2$. Radiat. Res. 1975,62(3):438-455.

[125] J.D. Buchanan, D.A. Armstrong, The Radiolysis of glyceraldehyde-3-phosphate dehydrogenase. Inter. J. Radiat. Biol. 1978,33(5):409-418.

[126] G.M. Gaucher, B.L. Mainman, G.P. Thompson, D.A. Armstrong, The $^{60}$Co gamma-radiolysis of aqueous papain solutions: repair and protection by cysteine. Radiat. Res. 1971,46(3):457-475.

[127] H. Abdoul-Carime, S. Cecchini, L. Sanche, Alternation of protein structure induced by low-energy (<18eV) electrons. Ⅰ.The peptide and disulfide bridges. Radiat. Res. 2002,158:23-31.

[128] J.M. Gebicki, J. Du, J. Collins, H. Tweeddale, Peroxidation of proteins and lipids in suspensions of liposomes, in blood serum, and in mouse myeloma cells. Acta. Biochim. Pol. 2000, 47(4):901-911.

[129] R.T. Dean, S. Fu, R.Stocker, M.J. Davies, Biochemistry and pathology of radical-mediated protein oxidation. Biochem. J. 1997, 324:1-18.

[130] H.A. Headlam, M.J. Davies, Beta-scission of side-chain alkoxyl radicals on peptides and proteins results in the loss of side-chains as aldehydes and ketones. Free. Rad. Biol. Med. 2002,32:1171-1184.

[131] H.A. Headlam, M.J. Davies, Markers of protein oxidation: different oxidants give rise to variable yields of bound and released carbonyl products. Free. Rad. Biol. Med. 2004,36(9):1175-1184.

[132] S. Uehara, H. Nikjoo, Track structure for low energy ions including charge exchange processes. Radiat. Prot. Dos. 2002, 99:53-55.

[133] B. Boudaïffa, P. CLoutier, D. Hunting, M.A. Huels, L. Sanche, Resonant formation of DNA strand breaks by low-energy (3 to 20 eV) electrons. Science 2000, 287:1658-1660.

[134] X. Pan, C. Cloutier, D. Hunting, L. Sanche, Dissociation electron attachment to DNA. Phys. Rev. Lett. 2003, 90:208102.

[135] B.D. Michael, P. O'Neil, A sting in the tail of electron tracks. Science, 2000, 287:1603-1604.

[136] F. Martin, P.D. Burrow, Z. Cai, P. Cloutier, D. Hunting, L. Sanche, DNA strand breaks induced by 0-4 eV electrons: the role of shape resonances. Phys. Rev. Lett. 2004, 93:068101.

[137] M.A. Huels, I. Hahndorf, E. Illenberger, L. Sanche, Resonant dissociation of DNA bases by subionization electrons. J. Chem. Phys. 1998, 108:1309-1312.

[138] H. Abdoul-Carime, M.A. Huels, F. Brüning, E. Illenberger, L. Sanche, Dissociative electron attachment to gas-phase 5-bromouracil. J. Chem. Phys. 2000, 113:2517-2521.

[139] G. Hanel, B. Gstir, S. Denifl, P. Scheier, M. Probst, B. Farizon, M. Farizon, E. Illenberger, T.D. Märk, Electron attachment to uracil: effective destruction at subexcitation energies. Phys. Rev. Lett. 2003, 90:188104.

[140] S. Denifl, S. Ptasinska, M. Cingel, S. Matejcik, P. Scheier, T. Märk, Electron attachment to the DNA bases thymine and cytosine. Chem. Phys. Lett. 2003, 377:74-80.

[141] H. Abdoul-Carime, S. Gohlke, E. Illenberger, Site-Specific Dissociation of DNA Bases by Slow Electrons at Early Stages of Irradiation. Phys. Rev. Lett. 92 (2004) 168103.

[142] R. Abouaf, J. Pommier, H. Dunet, Negative ions in thymine and 5-bromouracil produced by low energy electrons. Int. J. Mass Spectrom. 2003, 226:397-403.

[143] H. Abdoul-Carime, M. A. Huels, E. Illenberger, L. Sanche, Sensitizing DNA to Secondary Electron Damage: Resonant Formation of Oxidative Radicals from 5-Halouracils. J. Am. Chem. Soc. 2001, 123:5354-5355.

[144] I. Baccarelli, F. A. Gianturco, A. Grandi, N. Sanna, R. R. Lucchese, I. Bald, J. Kopyra, E. Illenberger, Selective Bond Breaking in $\beta$-D-Ribose by Gas-Phase Electron Attachment around 8 eV. J. Am. Chem. Soc. 2007, 129:6269-6277.

[145] R. Barrios, P. Skurski, J. Simons, Mechanism for Damage to DNA by Low-Energy Electrons. J. Phys. Chem. B 2002, 106:7991-7994.

[146] D. Antic, L. Parenteau, M. Lepage, L. Sanche, Low-Energy Electron Damage to Condensed-Phase Deoxyribose Analogues Investigated by Electron Stimulated Desorption of H$^-$ and Electron Energy Loss Spectroscopy. J. Phys. Chem. B 1999, 103:6611-6619.

[147] S. Ptasińska, S. Denifl, P. Scheier, T.D. Märk, Inelastic electron interaction (attachment/ionization) with deoxyribose. J. Chem. Phys. 2004, 120:8505-8511.

[148] Z.Cai, M. Dextraze, P. Cloutier, D. Hunting, L. Sanche, Induction of strand breaks by low-energy electrons (8–68 eV) in a self-assembled monolayer of oligonucleotides: Effective cross sections and attenuation lengths. J. Chem. Phys. 2006, 124:024705.

[149] M.H. du Penhoat, M.A. Huels, P Cloutier, J. Jay-Gerin, L Sanche, Electron stimulated desorption of H$^-$ from thin films of thymine and uracil. J. Chem. Phys. 2001, 114:5755-5764.

[150] C. Desfrançois, H. Abdoul-Carime, J.P. Schermann, Electron attachment to isolated nucleic acid bases. J. Chem. Phys.1996, 104:7792-7794.

[151] J. Berdys, I. Anusiewicz, P. Skurski, J. Simons, Damage to Model DNA Fragments from Very Low-Energy (<1 eV) Electrons. J. Am. Chem. Soc. 2004, 126, 6441-6447.

[152] S. Gohlke, A. Rosa, E. Illenberger, F. Brüning M.A. Huels, Formation of anion fragments from gas-phase glycine by low energy (0–15 eV) electron impact. J. Chem. Phys. 2002, 116:10164-10169.